\documentclass[12pt,preprint]{aastex}

\newcommand{\Ms}{\ensuremath{M_{\odot}}}
\newcommand{\eg}{{\it e.g.}}
\newcommand{\cf}{{\it c.f. }}
\newcommand{\ie}{{\it i.e.}}

\slugcomment{To appear in The Astrophysical Journal Letters}

\shorttitle{Dark star clusters}
\shortauthors{Banerjee \& Kroupa}

\begin{document}

\title{A new type of compact stellar population: dark star clusters}

\author{Sambaran Banerjee and Pavel Kroupa}
\affil{Argelander-Institut f\"ur Astronomie, Auf dem H\"ugel 71, D-53121, Bonn, Germany}

\email{sambaran@astro.uni-bonn.de, pavel@astro.uni-bonn.de}

\begin{abstract}
Among the most explored directions in the study of dense stellar systems is
the investigation of the effects of the retention of supernova remnants,
especially that of the massive stellar remnant
black holes (BH), in star clusters. By virtue of their eventual high central concentration,
these stellar mass BHs potentially
invoke a wide variety of physical phenomena, the most important ones being
emission of gravitational waves (GW), formation of X-ray binaries and
modification of the dynamical evolution of the cluster.
Here we propose, for the first time, that rapid removal of stars from the outer parts of
a cluster by the strong tidal field in the inner region of our Galaxy can unveil its BH sub-cluster,
which appears like a star cluster that is gravitationally
bound by an invisible mass. We study the formation and properties of such systems
through direct N-body computations and estimate that they
can be present in significant numbers in the inner region of the Milky Way.
We call such objects ``dark star clusters'' (DSCs) as they appear dimmer than normal star clusters of
similar mass and they comprise a predicted, new class of entities.
The finding of DSCs will robustly cross-check BH-retention;
they will not only constrain the uncertain natal kicks of BHs, thereby
the widely-debated theoretical models of BH-formation, but will also
pin-point star clusters as potential sites for GW emission for
forthcoming ground-based detectors such as the ``Advanced LIGO''.
Finally, we also discuss the relevance of DSCs for the nature of IRS 13E. 
\end{abstract}


\section{Introduction}\label{intro}

Compact remnants of massive stars in star clusters, which are neutron stars (NS)
and black holes (BH), form a dynamically interesting sub-population due to
their tendency of segregating towards the cluster's center and
augmenting their population density therein. In this respect,
the BHs are special in that they undergo a ``runaway'' mass segregation. 
These remnant BHs are typically several 10s of $\Ms$ heavy, enough to form a
Spitzer-unstable sub-system, provided a significant number of them are retained in
their parent cluster. Due to this instability (also called the mass-stratification
instability, \citealt{spz}), the continually sinking BHs cannot come to 
an energy equipartition with the local surrounding stars and finally end-up in
a central, highly concentrated sub-cluster made purely of BHs, which is self-gravitating and dynamically nearly 
isolated from the rest of the stellar cluster \citep{mert2004,mak2008,bbk2010}.

Such a dense environment of BHs is dynamically very active
due to the formation of BH-BH binaries via
3-body encounters \citep{hh2003} and their hardening by super-elastic 
encounters \citep{h75} with their surrounding BHs.
Studies of the dynamics of pure BH sub-clusters
using Monte-Carlo and direct N-body integration methods indicate that the dynamical BH-BH merger events they generate
are likely to contribute a significant gravitational wave (GW)
detection rate to the future ``Advanced LIGO'' (AdLIGO) and
``LISA'' GW observatories \citep{pzm2000,bcq2002,olr2006,moods2009,bbk2010,dng2011}. Such studies show that
a BH sub-cluster is typically self-depleted in a few Gyr due to the super-elastic dynamical encounters
and the resulting escape of the BHs \citep{olr2006,bbk2010}. The energy extracted
from the tight BH-BH binaries heats-up and expands the cluster's core \citep{mert2004,mak2008}, which can
be detectable by the future optical missions such as the ``GAIA'' mission.
Furthermore, the BHs can be
important for dynamically formed BH X-ray sources due to their encounters with
the surrounding stars \citep{ivn2010}. 
X-ray observations have indicated the presence of BH X-ray binary candidates in GCs
\citep{mcr2007,brs2010}.
The presence of a BH sub-cluster within a star cluster, therefore, has the potential to give
rise to a plethora of physical phenomena, all of which have significance to
upcoming prime missions such as the GAIA, AdLIGO and the present and future X-ray missions.

Is it possible to obtain any direct observational signature of the presence of a BH sub-cluster within
a star cluster?
We predict here, for the first time, that within a few kpc from the Galactic center,
rapid tidal stripping of
star clusters by the strong tidal field can expose its BH sub-cluster.
This would happen when the timescale of the preferential removal of
stars from the outer regions of the cluster is shorter or comparable to the encounter-driven self-depletion
timescale of its central BH sub-cluster (see above). Such a dissolved phase of the cluster would consist of a few
stars orbiting around a cluster of BHs and would observationally appear as a highly super-virial
star cluster with a large mass-to-light-ratio.

As we discuss here, a number of direct N-body computations
of model star clusters indeed support the formation of
such systems. These objects comprise a predicted, new class of
compact stellar populations which we name ``dark star clusters''.
The importance of dark star clusters
(hereafter DSC) is twofold: on one hand, if they are found to exist, then they guarantee
that star clusters are potential sites for
GW emission and formation of BH X-ray binaries and on the other hand, they naturally
constrain the uncertain natal kicks of BHs \citep{will2005}, as DSCs can form only if a significant number of BHs retain
in the cluster following their progenitor supernovae. This, in turn, restricts the theoretical models
of core collapse supernovae \citep{janka2007}.

\section{Computations}\label{comp}

We compute the evolution of model star clusters subjected to the Galactic tidal
field using the direct N-body integration method.
For our purposes, we use the state-of-the-art N-body integration code ``NBODY6'' \citep{ar2003}, which,
apart from utilizing a highly sophisticated numerical integration scheme \citep{ma92}, also
follows the evolution of the individual stars until their
remnant phases, using an analytic but well-tested stellar evolution recipe \citep{hur2000}. A unique
feature of NBODY6 is its use of highly accurate regularization methods
in resolving close encounters \citep{ks65,ar2003}. Furthermore, the code exploits the remarkable
hardware-accelerated computing capacity of Graphical Processing Units (GPUs) in integrating the
centers of masses. NBODY6 currently incorporates general-relativistic effects only
through analytic prescriptions of GW energy-loss.

\subsection{Dark star clusters}\label{darkcl}

We follow the evolution of initial
Plummer clusters \citep{pk2008} of single stars,
having masses between $10^4\Ms\lesssim M_{cl}(0)\lesssim7.5\times10^4\Ms$
and half-mass radii between $1.0{\rm~pc}\lesssim r_h(0)\lesssim3.5{\rm~pc}$.
All the clusters initially consist of zero-age-main-sequence (ZAMS)
stars with their masses $m$ chosen from the canonical initial mass function (IMF; \citealt{krp2001})
$\rho(m)\propto m^{\alpha}$, where $\alpha=-1.3$ for $0.07\Ms<m<0.5\Ms$
and $\alpha=-2.3$ (Salpeter index) for $m>0.5\Ms$.
Their metallicities are chosen to be solar, as suitable for our Galaxy's disk.
We assume for now that all the supernova remnants (\ie, BHs and NSs) receive
low natal kicks in general such that they remain bound to their parent clusters at their formation. 
Such models follow circular orbit around a point mass
of $M_c=2\times10^{10}\Ms$ representing the Milky Way bulge.

\begin{figure}
\centering
\vspace{-0.8 cm}
\includegraphics[width=11.5cm, height=7.0cm, angle=0]{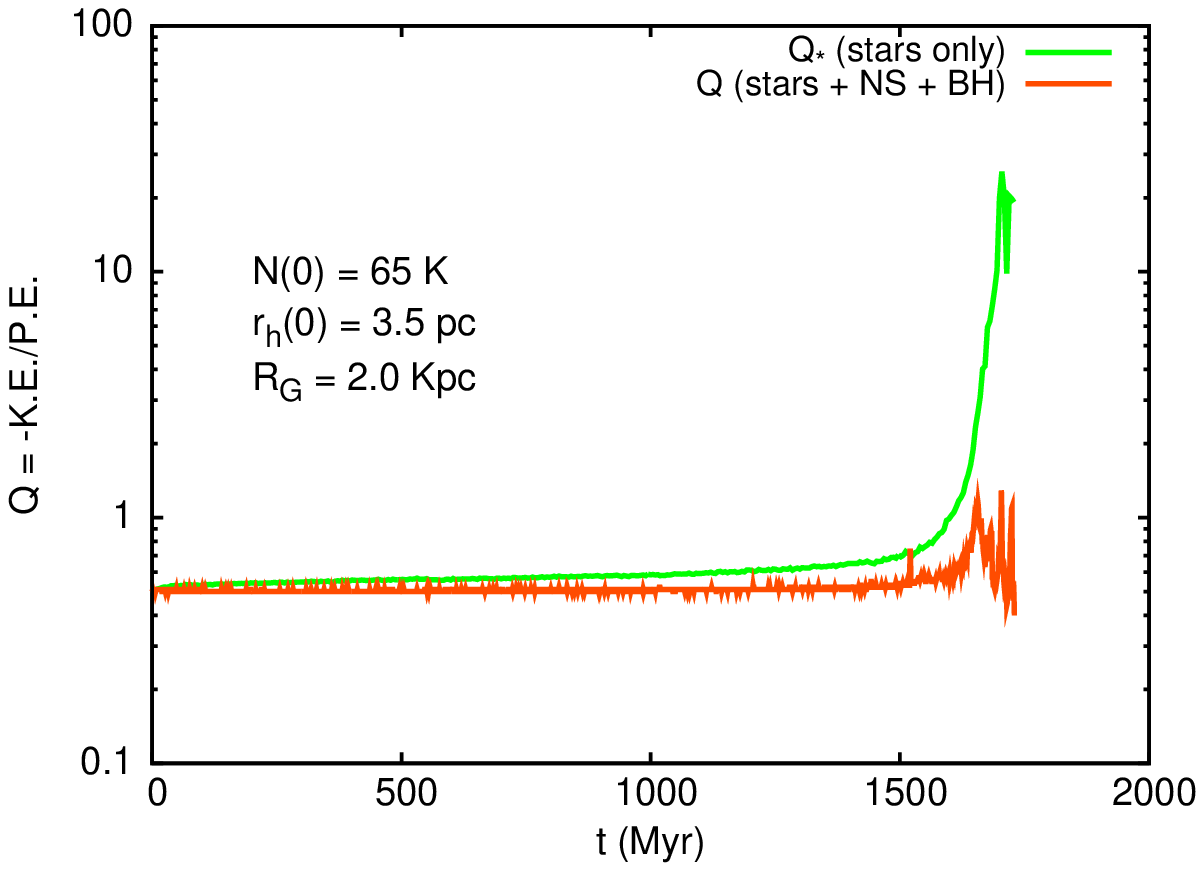}
\includegraphics[width=11.5cm, height=7.0cm, angle=0]{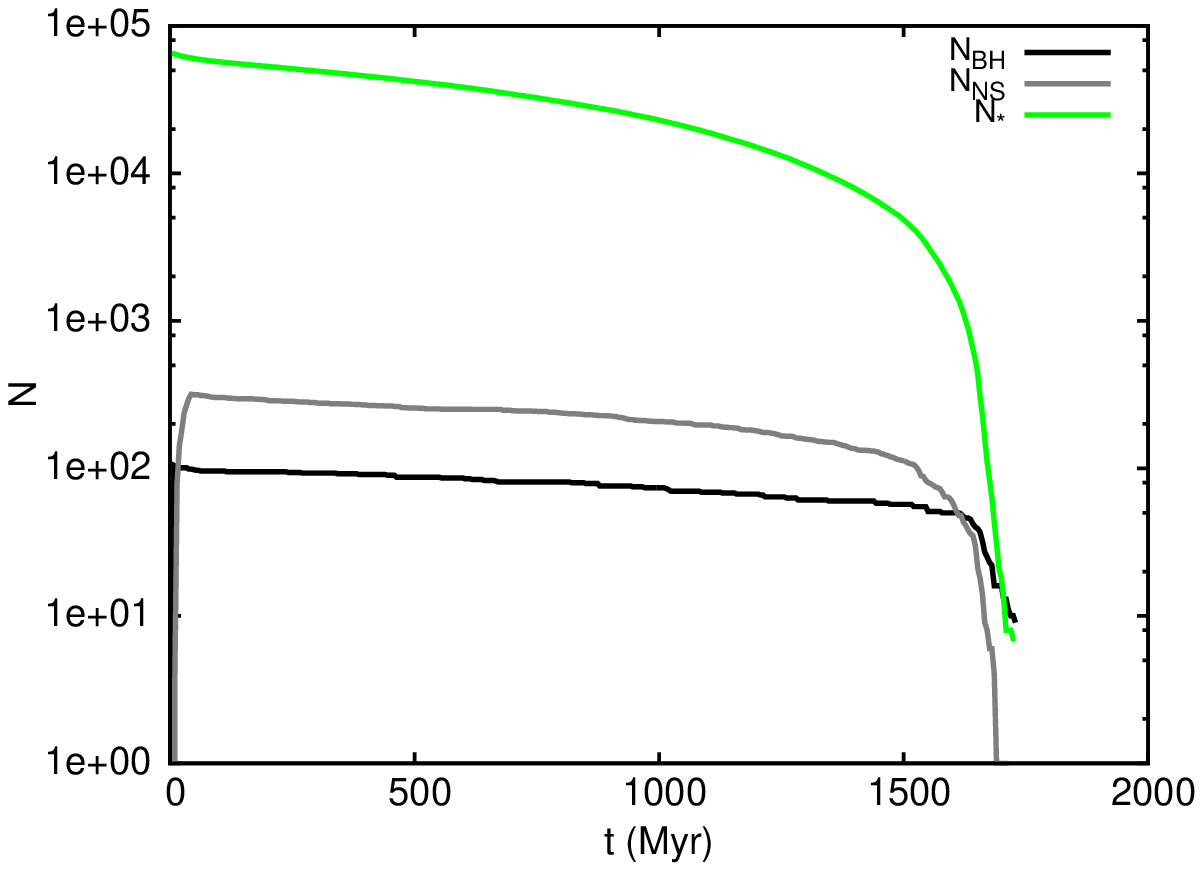}
\caption{{\bf Top:} Evolution of the virial coefficient of a Plummer cluster with
$N(0) = 6.5 \times 10^4$, $r_h(0)=3.5$ pc and $R_G=2.0$ kpc.
The lower (orange) curve represents the virial coefficient, $Q$,
for the whole cluster, \ie, including the luminous stars, NSs and BHs bound
to the cluster. The upper (green) curve represents the virial coefficient, $Q_\ast$, of the luminous
objects exclusively. The monotonic growth of $Q_\ast$ implies the approach
towards an apparent highly super-virial state --- the ``dark star cluster'' phase.
{\bf Bottom:} The corresponding evolution of the numbers of the luminous stars, $N_\ast$,
the NSs, $N_{\rm NS}$ and of the BHs, $N_{\rm BH}$.}
\label{fig:qevol}
\end{figure}

Fig.~\ref{fig:qevol} (top panel) shows an example of the evolution of the virial coefficient
for one of our computed model clusters with initially
$N(0)=6.5\times10^4$ stars and $r_h(0)=3.5$ pc, located at $R_G=2.0$ kpc Galactocentric distance.
The orange curve shows the time-evolution of the virial coefficient, $Q$,
for all the members bound to the cluster including the BHs and the NSs,
which mostly remains constant at $Q\approx0.5$, as it should be for the quasi-static relaxation
of a self-gravitating system through two-body encounters \citep{hh2003}.

The green curve in Fig.~\ref{fig:qevol} (top) shows the time-evolution of the virial coefficient
taking into account only the luminous objects, \ie, the nuclear burning stars and the white dwarfs (hereafter WD),
which are those an observer sees.
The BHs formed have masses $\approx10\Ms$ and the NSs are typically of $\approx2\Ms$, which are significantly more massive than
the majority of the remaining luminous members and hence are segregated to the cluster's center.
As the lower mass luminous stars in the outer regions are stripped by the external field,
the gravitational potential of these central invisible remnants becomes increasingly important. Therefore, 
the kinetic energy of the luminous sub-system increasingly exceeds the corresponding 
self-equilibrium (or quasi-static) value as the constituents perceive a potential that becomes increasingly deeper
than their self-potential. This correspondingly raises their exclusive virial coefficient, $Q_\ast$, above 0.5 as
in Fig.~\ref{fig:qevol} (top panel, green curve). The cluster thus evolves to a DSC state (see Sec.~\ref{dcpop}).
Note that while $Q_\ast$ reaches a very
high value, the system as a whole remains bound since $Q<1$ throughout (except at the
final dissolved state, not shown in the figure).

Notably, NSs form a few factors more often than the BHs and contribute significantly in elevating $Q_\ast$.
However, the NSs being lighter than the BHs, their sub-population occupies a more extended zone in the cluster's center.
This causes them to get stripped earlier than the BHs
(\cf Fig.~\ref{fig:qevol}, bottom panel) so that in the late evolutionary phase,
typically when $Q_\ast>1$, it is mostly the BH population that contributes to the augmented $Q_\ast$, thereby determining
the lifetime of the DSC phase (see Sec.~\ref{dcpop}).
Nevertheless, at large enough distances from the Galactic center, where slower tidal stripping
causes the DSC phase to appear later than the self-depletion of the
BH sub-cluster (see Sec.~\ref{intro}; \citealt{sighq93,mert2004,olr2006,moods2009,dng2010,bbk2010}),
the NSs constitute the primary dark component of the corresponding
DSC state (see Sec.~\ref{dcpop}). This self-depletion process, although operative for both the
NS and the BH sub-populations, is more efficient for the latter as it is significantly more concentrated.

\subsubsection{Galactic population of DSCs}\label{dcpop}

\begin{figure}
\centering
\includegraphics[width=11.5cm, height=8.0cm, angle=0]{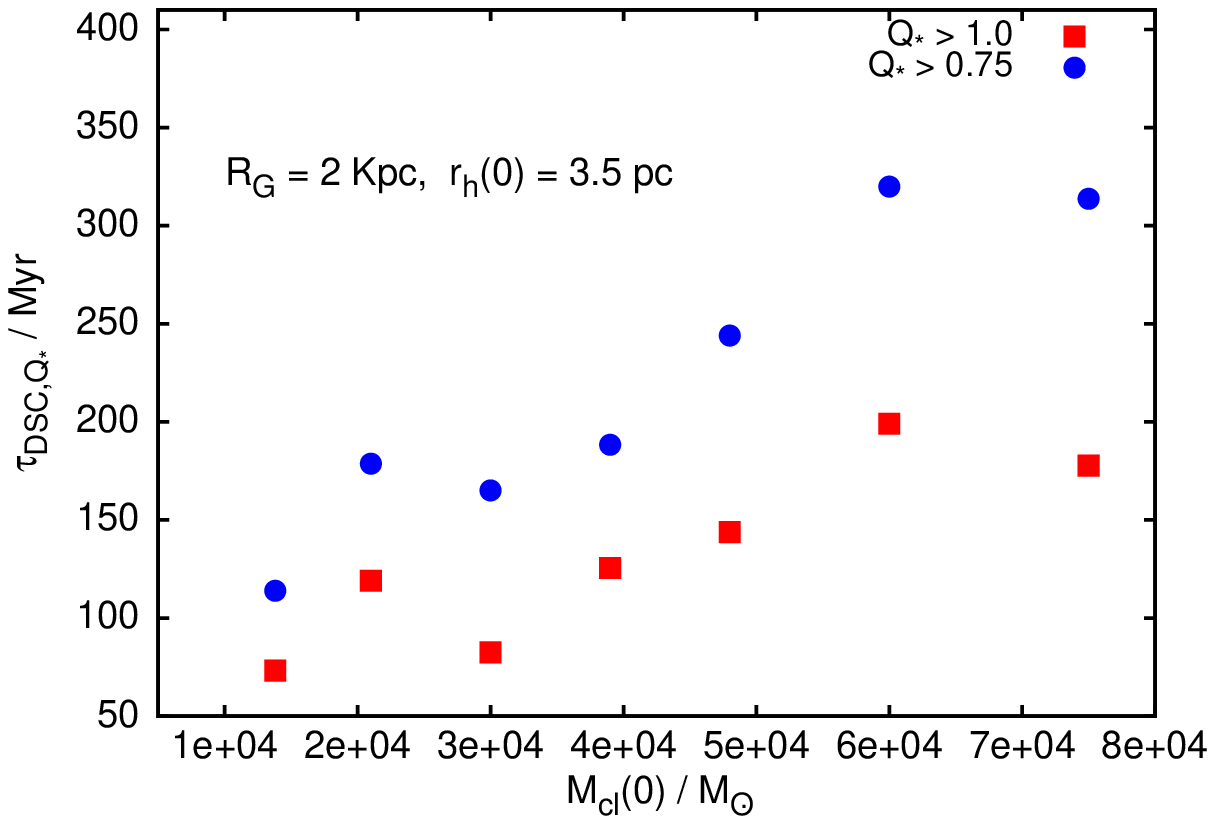}
\includegraphics[width=11.5cm, height=8.0cm, angle=0]{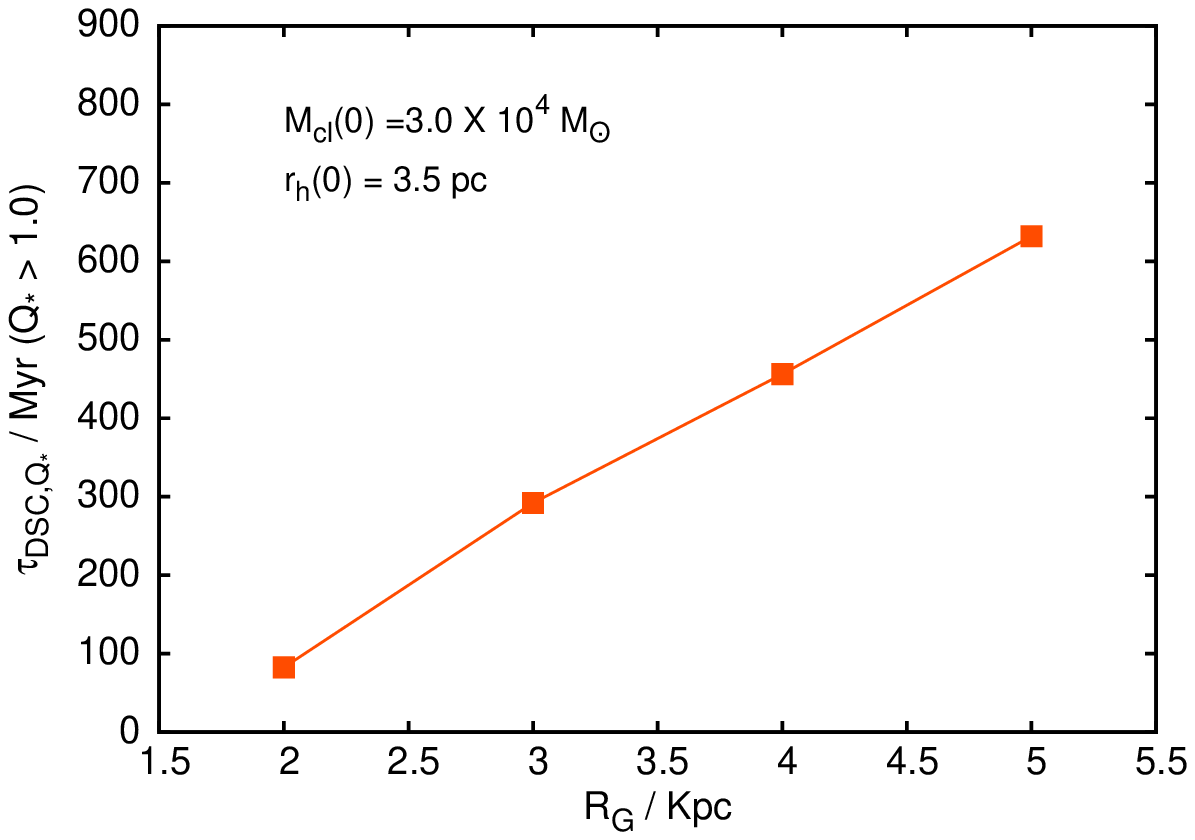}
\caption{{\bf Top:} The lifetimes, $\tau_{{\rm DSC},Q_\ast}$, of the DSCs with
increasing initial Plummer cluster mass $M_{cl}(0)$ ($R_G=2$ kpc, $r_h(0)=3.5$ pc).
{\bf Bottom:} $\tau_{{\rm DSC},Q_\ast}$
($Q_\ast > 1$) as a function of $R_G$ for a Plummer cluster with
$M_{cl}(0)=3\times10^4\Ms$ and $r_h(0)=3.5{\rm~pc}$.}
\label{fig:tfram}
\end{figure}

Fig.~\ref{fig:tfram} (top) shows the expected increasing trend in the lifetime of the DSCs
with initial cluster mass $M_{cl}(0)$ ($R_G=2{\rm~kpc}$, $r_h(0)=3.5{\rm~pc}$). The
DSC phase can be defined when the cluster appears unbound, \ie, $Q_\ast > 1.0$ or when
it appears significantly super-virial, which we take when $Q_\ast > 0.75$, and denote the
corresponding lifetimes by $\tau_{{\rm DSC},Q_\ast}$ with $Q_\ast=1.0$ and 0.75 respectively.
Fig.~\ref{fig:tfram} (bottom) shows $\tau_{{\rm DSC},1.0}$ against 
$R_G$ ($M_{cl}(0)=3\times10^4\Ms$, $r_h(0)=3.5{\rm~pc}$).
$\tau_{{\rm DSC},Q_\ast}$ also increases with increasing $R_G$ since the DSC takes a longer time to get depleted
in a weaker external field. Notably, for $R_G\gtrsim4$ kpc, the DSC state becomes NS-dominated.
Beyond $R_G\gtrsim5.5$ kpc, it takes more than the age of our Galaxy ($\approx10$ Gyr)
for the representative cluster to evolve to its DSC phase (\ie, $Q_\ast > 0.75$).

It then follows that for the present-day Galaxy the DSCs are formed from clusters with $M_{cl}(0)\gtrsim10^4\Ms$
within $R_G\lesssim5$ kpc, to be taken as representative numbers, and have lifetimes
$\tau_{{\rm DSC},1.0}\approx150$ Myr and $\tau_{{\rm DSC},0.75}\approx250$ Myr as conservative estimates.
Although these estimates are based on a point-mass tidal field, note that the DSCs' progenitor
clusters form in the Galactic disk and orbit on nearly circular paths (in the equatorial plane)
and hence would experience the same external field with an axisymmetric disk-like distribution of the same mass. The
$\tau_{{\rm DSC},Q_\ast}-R_G$ dependence can however be moderately modified as the clusters see a mass increasing moderately 
with $R_G$.

To estimate the Galactic population of DSCs,
we take the average star cluster formation rate (hereafter CFR) over the entire Galactic disk
($R_G < 10$ kpc) to be $0.16\Ms{\rm ~yr}^{-1}$ \citep{lrs2009} following a
Schechter initial cluster mass function \citep{scht76} over the mass range
$100\Ms \leq M_{cl}(0) \leq 2 \times 10^5\Ms$ \citep{lrs2009} which is
assumed to remain invariant over the last few hundred Myr.
This implies $\approx0.5$ clusters form per Myr which have properties ($M_{cl}(0)>10^4\Ms$ and $R_G<5$ kpc)
that must have them evolve to DSCs if a sufficiently high fraction of BHs are retained.
Assuming a steady state conversion to the DSC phase with
lifetime $\tau_{{\rm DSC},1.0}\approx150$ Myr (see above), the expected number
of $Q_\ast>1$ DSCs in the Galaxy within $R_G<5$ kpc is $N_{{\rm DSC},1.0}\approx75$.
For DSCs with $Q_\ast>0.75$, the corresponding number is $N_{{\rm DSC},0.75}\approx125$. Hence, a significant number of
DSCs can be expected in the inner Galactic zone.

\subsection{Can IRS 13E be a DSC?}\label{irs13E}

There has been recent concern with the widely debated IRS 13E (hereafter IRS13E); an extremely compact stellar
association of a few young, massive stars at a close projection to the Galactic center that apparently survives
the extreme tidal field by being bound by an invisible mass \citep{mail2004}. While a 
$\approx1300\Ms$ intermediate mass black hole (IMBH) was widely believed to be this invisible mass
\citep{pzm2002,mail2004,pz2006}, this possibility has recently been ruled out with a significant confidence by \citet{fritz2010}
through their newer proper motion measurements, leaving the nature of the invisible component of IRS13E
currently ambiguous. This status of IRS13E prompts us to consider whether its dark component can be an
ensemble of stellar-mass BHs instead. To that end, we perform preliminary N-body calculations to determine
whether a DSC configuration resembling IRS13E can be a possible fate of a star cluster very close to the
Galactic center.

These computed clusters follow circular orbits around the Galactic SMBH
(a central mass of $\approx5.4\times10^6\Ms$ which is a combination of the mass of the SMBH and
that of the nuclear star cluster within $R_G\lesssim1$ pc) within $R_G<$ few pc,
where star formation has been shown to possibly lead to a top-heavy stellar IMF
\citep{bko2009,ns2005,nayak2006,bonn2008}.
Since we are primarily interested in the final state, we begin the N-body calculations from an evolved
phase of the cluster, for computational ease. Therefore, we initiate the computations with Plummer clusters
made of stars which are pre-evolved until $\tau_0\approx3.5$ Myr age. This age
is slightly earlier than when the most massive star (of $\approx 150\Ms$) evolves to a BH.
All the stars are thus still on their main sequence.

Observations indicate a very flat IMF ($\alpha=-0.45\pm0.3$; see \citealt{bko2009}) for stars
close to the Galactic center and also a dearth of low-mass stars. The latter is evident from a declining density
of B-stars away from the central SMBH \citep{bko2009} and a significant lack of coronal X-ray emission \citep{ns2005}
from the SgrA$^\ast$ field. It is currently unclear from the literature what would be the lower-mass limit
of such an IMF, which, in turn, determines the $\alpha$ of the mass function (MF) and
its lower limit $M_l$ at cluster-age $\tau_0$.
In this preliminary study, we simply take the total number of stars in the cluster $N$,
$\alpha$ and $M_l$ at age $\tau_0$ as free parameters.
The upper MF limit is chosen to be the canonical $150\Ms$ and the stellar metallicity is solar.
These limits are the ZAMS values and they are appropriately reduced during the $\tau_0\approx3.5$ Myr
pre-evolution by the NBODY6's built-in stellar evolution prescription (see above). Like the computations in Sec.~\ref{darkcl},
we retain all the BHs (of $\sim 10\Ms$ as obtained from within NBODY6) in the cluster.
At the beginning of the computations, the models are taken compact to a similar extent as IRS13E \citep{mail2004},
with half-mass radii between $0.02{\rm~pc}<r_h(0)<0.04{\rm~pc}$ and they orbit within $2{\rm~pc}<R_G<4{\rm~pc}$.

\begin{figure}
\centering
\vspace{-1.0 cm}
\includegraphics[width=11.0cm, height=6.5cm, angle=0]{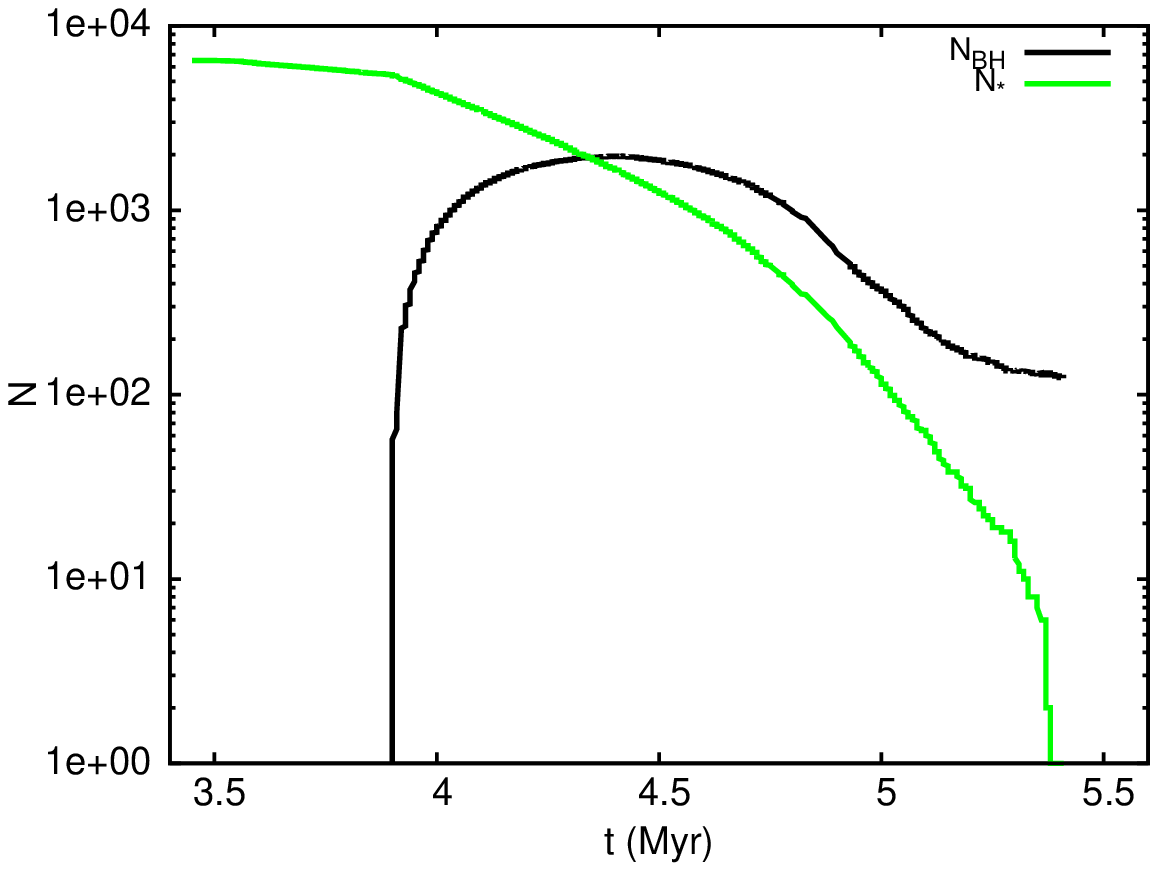}
\includegraphics[width=11.2cm, height=6.5cm, angle=0]{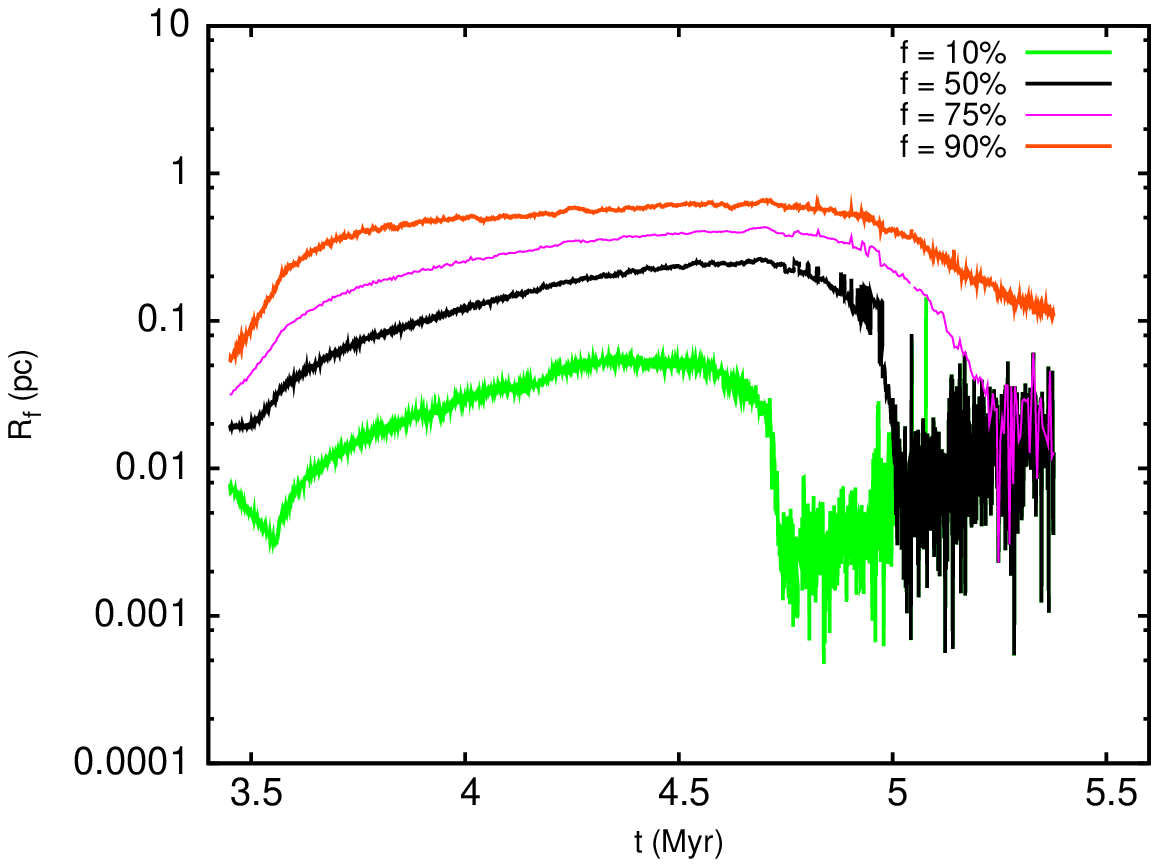}
\includegraphics[width=11.0cm, height=6.5cm, angle=0]{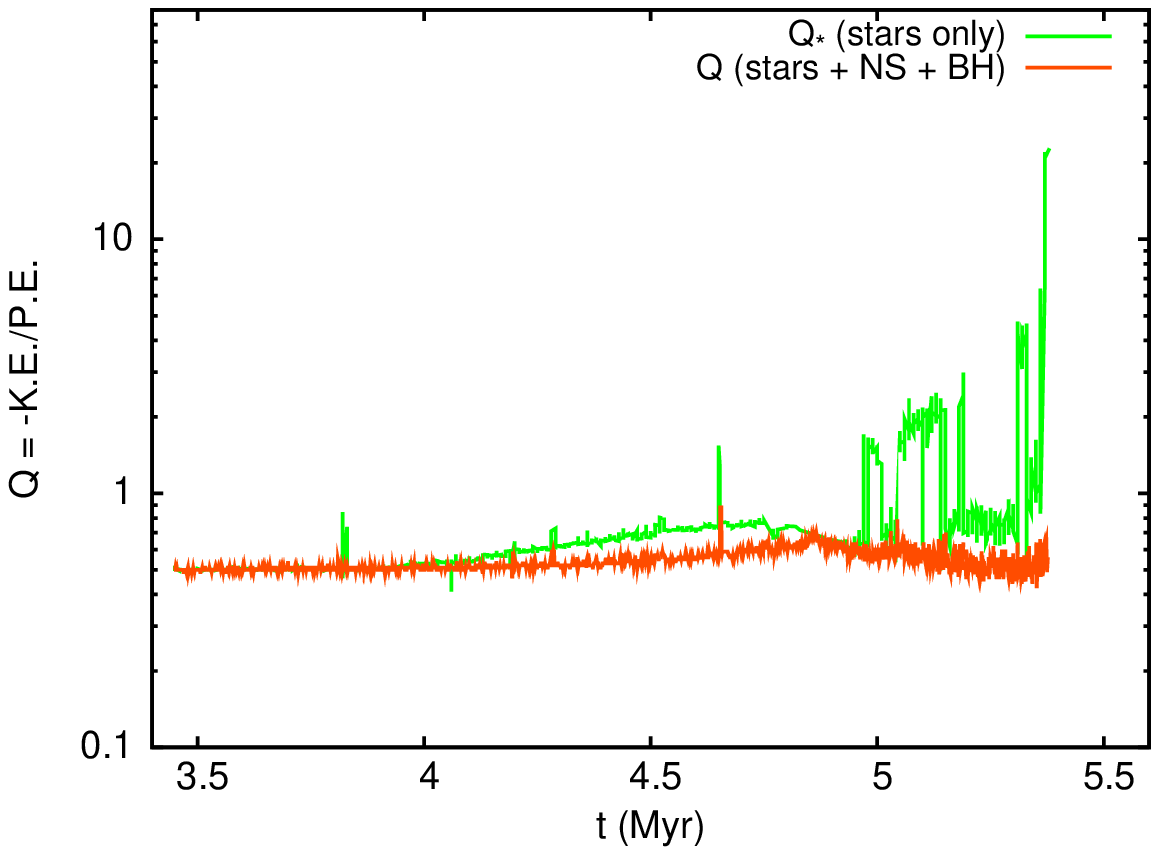}
\caption{The evolution of the numbers of the luminous members, $N_\ast$, and of the BHs, $N_{\rm BH}$, for a
model star cluster computation, leading to an IRS13E solution. The direct N-body computation
starts with a Plummer cluster of pre-evolved stars having $N(0)=6.5\times10^3$, $r_h(0)=0.025$ pc and $R_G=4$ pc,
where the stars are taken from a $\alpha=0$ MF within $35\Ms < M < 150\Ms$ ZAMS mass (see Sec.~\ref{irs13E}).
The evolution of the Lagrangian radii of the system is shown in the middle panel.
The final state of the cluster resembles that of IRS13E.
The evolution of the viral coefficient $Q$, for the whole system, and
$Q_\ast$, for the luminous sub-system is shown in the bottom panel.}
\label{fig:irs13}
\end{figure}

We find that to reach a IRS13E-like state, \ie,
a state where typically $\approx10$ luminous, young stars are very tightly bound to a cluster of $\approx100$
BHs, it takes $N\approx6.5\times10^3$ and a rather extreme lower cutoff of $M_l\approx35\Ms$
for MF index $\alpha\approx0$ (at age $\tau_0$), this $\alpha$ being
close to the upper limit of the \citet{bko2009} index. A steeper MF requires
even higher $M_l$.
As a demonstration, it can be seen in Fig.~\ref{fig:irs13} (top panel) that this cluster
eventually evolves to a configuration consisting of $\approx130$ BHs,
comprising a $\approx1300\Ms$ dark component, orbited by $<10$ young stars.
Although the half-mass radius of the system expands at the beginning of the evolution, primarily
driven by the mass loss through
the massive stars' winds and their supernovae, it finally collapses to $\approx0.02$ pc, as shown in
Fig.~\ref{fig:irs13} (middle panel), by the time the system arrives at the above configuration.
The luminous members in this state include 1-2 O-stars (those still on the main sequence) and a
few helium and Wolf-Rayet stars, thereby being of similar variety as observed in IRS13E.
The final state of the system, therefore, resembles IRS13E in terms of
compactness and stellar content \citep{mail2004}.
Fig.~\ref{fig:irs13} (lower panel) shows that the luminous sub-system becomes highly super-virial,
\ie, a DSC, as the IRS13E-like state occurs.

The calculations presented in this section
are preliminary which suggest an intermediate state of a cluster, close to the Galactic center,
that would evolve to an IRS13E-like configuration. It remains an open question whether
any reasonable initial cluster evolves to such an intermediate phase, which
depends on the low-mass limit
of the Galactic-central IMF, its index and the initial mass and compactness of the cluster, at a given $R_G$. A scan over
these parameters, beginning with much larger and more compact clusters, is necessary to determine such possible initial
configuration(s), which is much more compute-intensive. Although the above $M_l$ appears too high for a
lower cut-off, its progenitor cluster would have a smaller limit taking into account the rapid tidal
stripping over the 3.5 Myr pre-evolution. Tidal stripping also implies a steeper initial index, in better
agreement with the observed one.

Although we over-simplify by effectively ignoring the dynamical evolution of the cluster during the stellar
pre-evolution, the dynamical history is not instrumental in determining the occurrence of a IRS13E-type
state which happens merely due to the competition between stellar evolution and tidal dissolution of the cluster.
It is enough that the cluster remains bound by the time most of the stars become BHs quenching the wind mass loss;
the system would then core-collapse to a compact configuration irrespective of its history. This is in contrast
with the ``classical'' DSCs discussed in Sec.~\ref{darkcl} whose formation depends crucially on mass segregation.

A potential drawback of this ``in-situ'' model is that it is likely to significantly over-produce
the number of massive stars seen in the Galactic center.

Given these drawbacks, the calculations in this section are only to suggest that IRS13E might be a DSC but are by no means
conclusive.

\section{Discussion}\label{discuss}

Our calculations (Sec.~\ref{darkcl}) signify that a gravitationally bound star cluster
naturally evolves to an apparent super-virial state,
while remaining bound as a whole, as a consequence of the interplay between the dynamics and the evolution of stars,
provided a significant number of stellar remnants survive in the bound system after their formation via supernovae.
The existence of such dark star clusters is a first-time prediction and serves as an excellent cross-check
of the retention of supernova remnants in star clusters, the effects of which is widely explored.
The very presence of DSCs would kill two birds with one stone by having
consequences on the widely debated theoretical stellar collapse models \citep{janka2007} due to
the implied direct constraints on natal kicks, and by securing star clusters as potential sources
for the forthcoming ``Advanced LIGO'' \citep{hari2010} GW detector \citep{pzm2000,bbk2010}.
Given that the DSCs must have a significant population
in the inner region of our Galaxy, it is tempting to conduct
a survey of intermediate-aged (see below) stellar assemblies to identify them.

\begin{figure}
\centering
\includegraphics[width=11.5cm, height=7.5cm, angle=0]{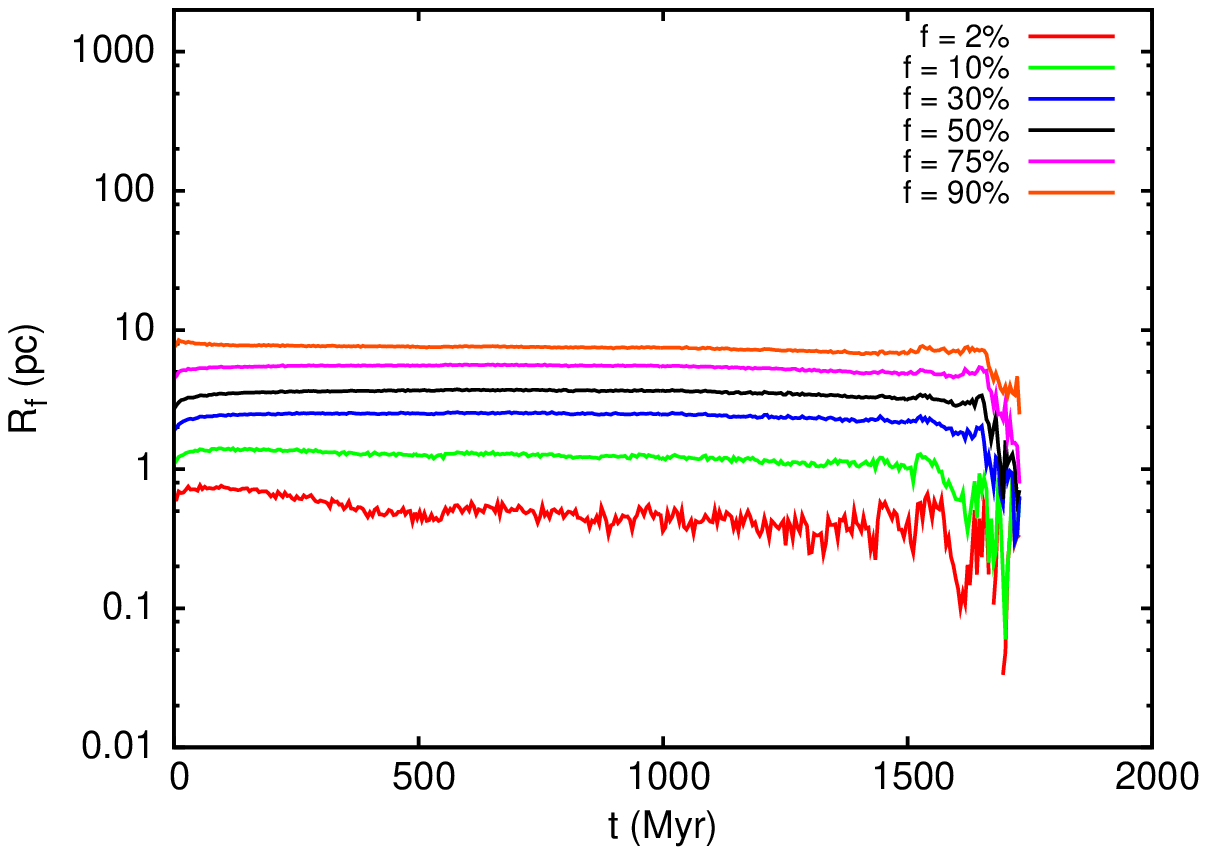}
\includegraphics[width=11.5cm, height=7.5cm, angle=0]{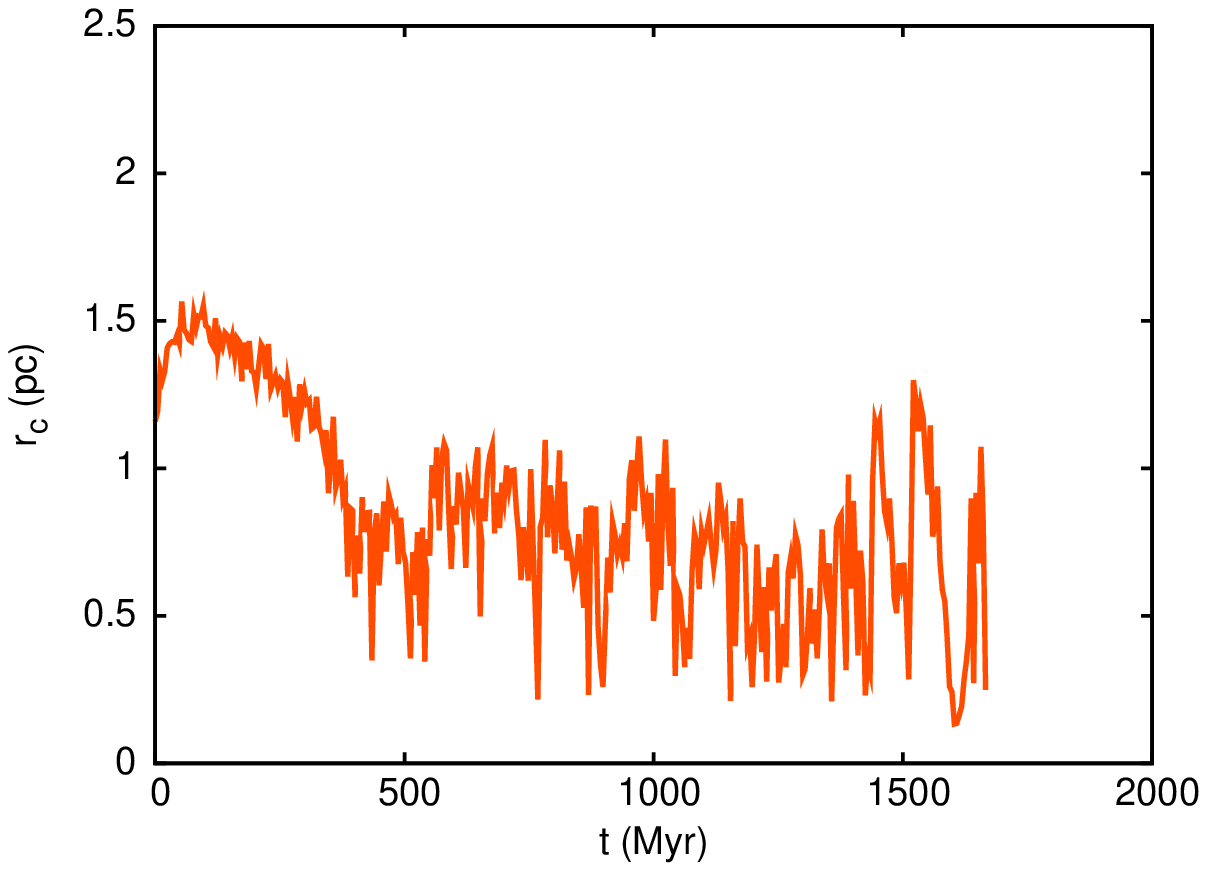}
\caption{The evolution of the Lagrange radii (top) and the core radius (bottom) of the example
cluster of Fig.~\ref{fig:qevol}. The half-mass radius (black curve, top panel) remains nearly constant
throughout and the core also remains compact following an initial brief contraction phase.}
\label{fig:lagrev}
\end{figure}

Dark star clusters can be observationally distinguished from actually dissolving clusters (which are also super virial
and hence expanding and are generally young) through their compact sizes in spite of their intermediate ages,
as shown in Fig.~\ref{fig:lagrev}.
That the apparent super-virial state of a DSC is not due to an IMBH, would be indicated by the absence
of a central cusp in its velocity-dispersion profile, which would be there otherwise \citep{noyola2008}.

The present studies justify DSCs as predicted, new type of compact stellar populations.
An immediate improvement over this study is to explore any effects of varying orbital
eccentricity and different initial profile types (\eg, using King instead of Plummer profiles).
Also, the effect of a varying star (and hence cluster) formation rate
over the Galactic disc needs to be incorporated to determine the predicted Galactic DSC population, as an
improvement over our assumption of a uniform average cluster formation. Another development would be
to consider a disk-like mass distribution instead of a central point-mass (but see Sec.~\ref{dcpop}).

The preliminary computations in Sec.~\ref{irs13E} suggest that the dark component of IRS13E can be an ensemble
of stellar mass BHs and IRS13E may therefore perhaps be a DSC. However, this conclusion should, for now, be taken as
being suggestive rather than conclusive due to the
drawbacks discussed (Sec.~\ref{irs13E}). We postpone a more detailed and self-consistent study
on this issue in a future paper.

\end{document}